# Impact of 3D curvature on the polarization orientation in non-Ising domain walls

*Ulises Acevedo-Salas[1], Boris Croes[1], Yide Zhang[1], Olivier Cregut[1], Kokou Dodzi Dorkenoo[1], Benjamin Kirbus[2], Ekta Singh[2], Henrik Beccard[2], Michael Rüsing[2], Lukas M. Eng[2,3], Riccardo Hertel[1], Eugene A. Eliseev[4], Anna N. Morozovska[5,+], Salia Cherifi-Hertel[1,\*]*

[1] Université de Strasbourg, CNRS, Institut de Physique et Chimie des Matériaux de Strasbourg, UMR 7504, 67034, Strasbourg, France

[2] Institute of Applied Physics, Technische Universität Dresden, 01062 Dresden, Germany

[3] ct.qmat: Würzburg-Dresden Cluster of Excellence - EXC 2147, TU Dresden, Germany

[4] Institute for Problems of Materials Science, National Academy of Sciences of Ukraine, Krjijanovskogo 3, 03142 Kyiv, Ukraine

[5] Institute of Physics, National Academy of Sciences of Ukraine, 46, pr. Nauky, 03028 Kyiv, Ukraine

Corresponding authors:
\* E-mail: salia.cherifi@ipcms.unistra.fr
+ E-mail: anna.n.morozovska@gmail.com

ABSTRACT:

Ferroelectric domain boundaries are quasi-two-dimensional functional interfaces with high prospects for nanoelectronic applications. Despite their reduced dimensionality, they can exhibit complex non-Ising polarization configurations and unexpected physical properties. Here, the impact of the three-dimensional (3D) curvature on the polarization profile of nominally uncharged 180° domain walls in $LiNbO_3$ is studied using second-harmonic generation microscopy and 3D polarimetry analysis. Correlations between the domain wall curvature and the variation of its internal polarization unfold in the form of modulations of the Néel-like character, which we attribute to the flexoelectric effect. While the Néel-like character originates mainly from the tilting of the domain wall, the internal polarization adjusts its orientation due to the synergetic upshot of dipolar and monopolar bound charges and their variation with the 3D curvature. Our results show that curved interfaces in solid crystals may offer a rich playground for tailoring nanoscale polar states.

KEYWORDS: *Ferroelectric domain walls, curvature, non-Ising domain walls, second-harmonic generation, phase-field simulations, lithium niobate.*

In ferroic materials, the sample geometry can significantly influence the order parameter because of boundary conditions and charge distributions at the surface. A correlation between the topology of the sample surface and the ferroic order occurs in many systems[1], which opens a smart pathway towards engineering specific materials properties by tailoring the surface morphology. In the case of ferromagnetic materials, such situations have been investigated in detail over the past years for several complex geometries, including rolled-up structures[2], hemispherical shells[3], and several shapes therealike[4]. In studies of such artificially structured materials, the role of surface curvature has become a topic of growing interest, as theory predicted that curvature might lead to a chiral symmetry breaking of the order parameter[5,6] and, as a result, to surprising effects like non-reciprocal wave propagation and asymmetric domain-wall dynamics[7,8]. While the topic of curvilinear magnetism appears to be in full bloom[9], identifying and understanding the role of curvature in ferroelectric materials is still in its infancy.

Owing to the recent progress in the fabrication of curvilinear crystalline ferroelectric oxides and the modeling of their properties, it has become clear that curvature can provide a new pathway towards the polarization control at surfaces[10], and for tailoring exotic polarization textures such as bubble domains[11,12]. Functionality enhancement and variation of the phase transition temperature mediated by curvature are evidenced in ferroelectric nanocylinders[13]. Dong *et al.*[14] demonstrated that the ultraflexibility of freestanding crystalline barium titanate membranes is directly connected to the dynamic variation of the ferroelectric domain structure with curvature. In a similar context, Zhou *et al.*[15] reported periodic domain patterns in a rippled $BaTiO_3$ freestanding film in which the ferroelectric domain structure is linked to the local curvature and can be controlled by applying a pressure through flexoelectric coupling.

Interfaces with irregular shapes in ferroelectric materials can also result in unexpected local physical properties. Recent studies have shown that curved interfaces can naturally develop in ferroelectric/dielectric heterostructures through epitaxial strain and strain gradients, resulting in complex three-dimensional (3D) domain structures[16], or periodic dipole waves[17] owing to the complex balance between elastic and electrostatic energies. The control of elastic effects in epitaxial heterostructures is, however, limited by the substrate clamping. The engineering of competing electrostatic and strain forces, which can lead to exotic polar topological structures, is better enabled by setting the epitaxial film free[18] through complex growth procedures[19].

Bent and crumpled homo-interfaces formed by domain boundaries are also expected to show enhanced functional properties. In fact, a ferroelectric domain wall can be regarded as a homo-interface that can be programmed, (re)moved, or bent quasi at will, simply by applying an external field. Zhang et al.[20] reported a p–n junction within a single domain wall resulting from the wall bending in $BiFeO_3$ thin films. Maksymovych et al.[21] established direct correlations between the metallic conductivity of domain walls and the conical shape of their adjacent ferroelectric domains in otherwise insulating $PbZrTiO_3$ thin films. Both theory[22,23] and experimental studies[24,25] confirm that the local charge and the related conductivity can be affected by the tilt angle of the walls with respect to the polar axis. Clear correlations between the domain wall conductivity and the variation of bound charges due to the tilt angle in lithium niobate crystals have been evidenced by combining 3D noncolinear Cherenkov SHG and electrical conductivity measurements[24,26,27]. Based on nanoscale studies combining focused-ion-beam-assisted scanning electron microscopy and tomographic piezo-response force microscopy, Roede et al.[28] showed the same effect in hexagonal manganites.

Despite the intensive research on the morphology-structure-functionality relation in ferroelectrics, the effect of curvature on the internal structure of ferroelectric domain walls has hitherto not been addressed. Experimental studies on the variation of the internal polarization as a function of the domain wall curvature might uncover new physical aspects on, e.g., the propagation of domain walls in disordered media[29] in which pinning[30] and roughening[31] are dominant, and it bears new prospects for next-generation domain wall-based nanotechnology[32–35].

In this work, the impact of curvature on the internal polarization of a nominally uncharged 180° ferroelectric domain walls is investigated in a z-cut $LiNbO_3$ bulk crystal using non-invasive 3D nonlinear optical microscopy and polarimetry analysis. The ferroelectric polarization profile of an "arc-shaped" domain wall segment is derived through colinear 3D SHG microscopy in back-reflection geometry and polarimetry analysis (Figure 1). The variation of the 3D SHG response with the light polarization is analyzed pixel-by-pixel throughout the domain wall profile. The analysis of the local second-harmonic emission and its polarization dependence[36] reveals an overall Néel-like internal structure that follows the mean curvature of the wall. Nevertheless, clear local deviations of the internal polarization with respect to the surface normal of the domain wall are observed. This result is discussed in terms of the synergetic action of monopolar bound charges due to the wall tilting, and dipolar charges due to the Néel-like character. The polarization arrangement at the homo-interface formed by a wall separating two ferroelectric domains is affected by the complex interplay between the electrostatic and elastic forces[37,38], in a similar way as in crystalline heterointerfaces. The balance of the different forces at play in non-Ising domain walls, namely electrostatic forces and flexoelectric coupling, enhances the sensitivity to morphology variations, thereby providing a new degree of freedom to control the polarization.

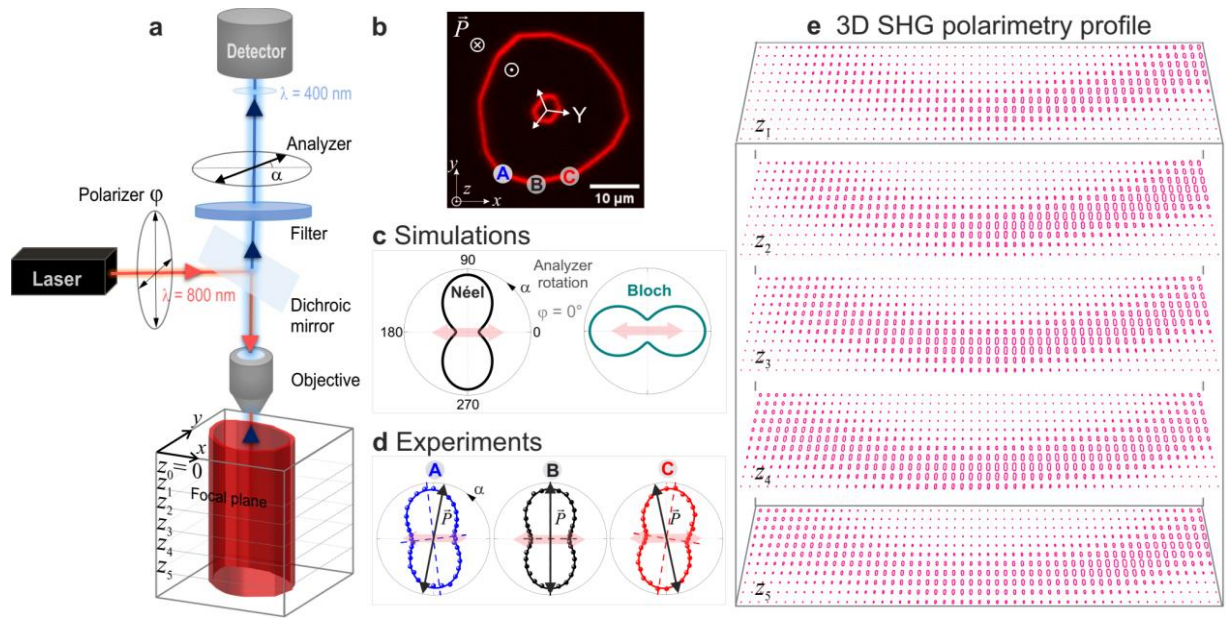

**Figure 1.** 3D SHG polarimetry analysis. (a) Schematic representation of the SHG microscope. (b) Isotropic SHG image displaying a rounded polygonal-shape c-domain in congruent Mg:LiNbO$_3$. A small concentric hexagonal domain with reversed polarity is used as a reference for the crystallographic orientation of the Y-axis with respect to the ($x,y,z$) laboratory coordinates system. (c) Simulated SHG polarimetry responses expected in the case of straight Néel and Bloch-type domain walls oriented parallel to $x$-axis. (d) SHG polarimetry response measured at three positions (A, B, C in panel b) along the "arc-shaped" domain wall segment at a depth of $z = 160$ μm below the surface. The experimental data (scattered dots) can be fitted very well by assuming a Néel model (continuous line). The presented simulations and experiments are obtained for a fundamental beam polarized along $x$-axis ($\varphi = 0°$), as depicted by the pink arrow. (e) Pixel-by-pixel 3D polarimetry mapping of the curved domain wall portion connecting points A,B,C at different depths $z$ ($z_1$=52.5 μm, $z_2$=73.5 μm, $z_3$=107.1 μm, $z_4$=136.5 μm, $z_5$=168 μm).

RESULTS AND DISCUSSION

Because of their trigonal symmetry and stoichiometry, uniaxial ferroelectrics such as LiNbO$_3$ or LiTaO$_3$ crystal families are expected to show rather simple domain structures of hexagonal or triangular shapes[39]. Nonetheless, a large variety of complex polygonal shapes can be obtained by non-equilibrium switching conditions due to incomplete or ineffective screening of the depolarization field during the poling procedure, as explained by Shur *et al.*[40]. This can be caused, *e.g.*, by a fast sweeping of the applied electric field, or by pyroelectric effects induced by a local laser heating. To create curved walls, the uniform polarization of a *z*-cut congruent Mg:LiNbO$_3$ crystal was locally switched under UV-laser irradiation, following the procedure detailed by Haussmann *et al.*[41]. We obtain a large domain of rounded shape using laser-assisted switching (see Figure 1b). This domain extends in the *z*-direction throughout the whole crystal thickness (200 µm), and it contains in its core a small conventional hexagon-shaped domain with reversed polarity. This small inner domain serves as a reference object for the crystal orientation since the characteristic *Y*-walls surrounding the hexagonal-shape domain are oriented along three equivalent crystallographic axes ($[11\bar{2}0], [2\bar{1}\bar{1}0], [1\bar{2}10]$)[42].

Owing to its quadratic dependence on the electric field and its high sensitivity to local symmetry breaking, nonlinear optical microscopy based on the second-harmonic process allows for the detection of nanostructures that are *a priori* incompatible with the resolution limit of optical methods[43]. In this context, SHG microscopy has recently emerged as a method of choice to probe the local ferroelectric polarization at domain walls[36,44–46]. The acquisition of the localized nonlinear optical emission at the domain walls allows a noninvasive 3D profiling of domain walls embedded in the volume of ferroelectric crystals[47–49], while the polarization analysis of the second-harmonic emission has proven its efficiency in detecting the local symmetry lowering at the domain wall regions[36]. This capability made it possible to unveil a non-Ising and chiral internal structure of the wall[44] with an overall in-plane polarization component along the domain wall (Bloch-like) or perpendicular to it (Néel-like). Local SHG polarimetry hence provides an unambiguous signature of the internal polarization structure of the domain wall as demonstrated by the complementarity of the polar plots displayed in Figure 1c for Néel and Bloch-like configurations.

Local 2D pixel-by-pixel polarimetry analysis of ferroelectric domains[50] and polar domain walls[51] was previously reported by Uesu's group. In the present study, we use colinear SHG microscopy with a 3D pixel-by-pixel polarimetry analysis to probe the variation of the internal polarization of a ferroelectric domain wall with its morphology throughout the sample volume. It is worth noting that the effect of the domain wall morphology is not limited to curvature. According to theoretical predictions, a change in the orientation of the wall with respect to the crystal axes is also expected to affect the internal structure and chiral character of ferroelectric domain walls[39,52]. To distinguish the effect of curvature from other geometrical effects, we consider here only small changes (on the order of ten degrees) in the orientation of the wall. In addition, we take great care in positioning the curved domain wall portion along the *x*-axis to avoid unwanted effects related to the beam focus which may result in distortions of the SHG polar plots, as discussed elsewhere[53].

Figure 1d displays local SHG polar plots measured at three different points labeled A, B, and C along the "arc-shaped" domain wall. These three points show similar SHG polar plots that can be fitted very well with a Néel model (continuous lines in Figure 1c) resulting from an overall polarization aligned perpendicular to the domain wall. This observation is further confirmed by the 3D variation of the SHG polar plots throughout the domain wall profile (Figure 1e). As a general trend, the internal polarization seems to follow the mean curvature of the domain wall, while preserving an overall Néel-like character (i.e., showing an overall in-plane polarization component perpendicular to the domain wall) throughout the sample volume.

Let us now examine the domain wall profile in terms of the local polarization orientation and morphology. Information on the morphology of the domain wall (Figure 2b) is obtained based on the 3D SHG image (Figure 2a), while the local orientation of the internal polarization angle $\theta_P$ is derived by fitting each polar plot, pixel-by-pixel, as explained in Supplementary note S2. By repeating the pixel-by-pixel fitting procedure at different focusing depths, ranging from 50 µm to 170 µm below the surface of the sample, we obtain a full 3D SHG analysis of the curved domain wall segment. The combination of the *x,y,z* coordinates of the wall morphology derived from the 3D SHG image (see Figure S2), and the angle $\theta_P$ obtained as an output parameter of the pixel-by-pixel fitting of the SHG polarimetry response, allows us to represent the internal polarization profile (Figure 2c). Selected results (fitting factor $R^2 > 0.96$) are presented in the form of bidirectional arrows. Comparing the polarization angle with the surface normal, i.e., $\theta_P$ (Figure 2c) versus $\theta_n$ (Figure 2d), demonstrates deviations up to ± 15° of the internal polarization with respect to the domain wall normal vector.

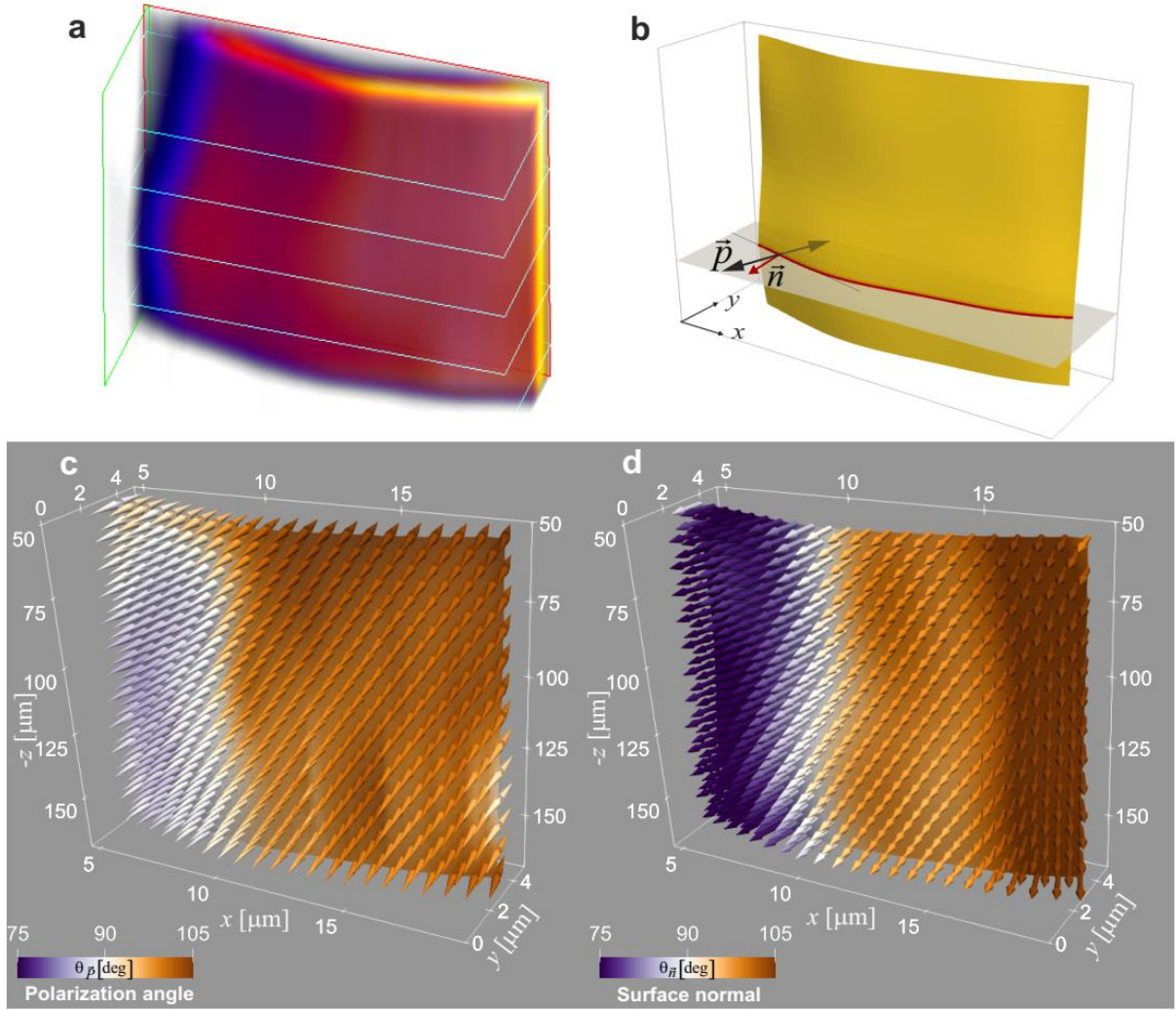

**Figure 2.** (a) 3D SHG image measured in a sample volume of 20 × 8 × 125 µm³ displaying the profile of a curved domain wall segment. The image was recorded at polarizer and analyzer angles φ = 0° and α = 90°, respectively. (b) Reconstruction of the wall morphology based on the experimental data illustrating the internal polarization vector $\vec{P}$ and the normal vector of the domain wall surface $\vec{n}$. (c) 3D polarization profile of the domain wall, derived from the pixel-by-pixel fitting of the 3D SHG polarimetry data as detailed in Supplementary note S2. As a comparison, the surface normal vector $\vec{n}$ is displayed in panel (d). The color map (purple-white-orange) in panels c-d represents the orientation angle of the vectors $\vec{n}$ and $\vec{P}$ taken with respect to the *x*-axis.

To better understand the origin of this deviation angle, we analyze the charge density distribution and study its variation with the domain wall curvature. Let us first consider the bound charge density distribution $\rho_{bound}$ induced at tilted domain walls with respect to the polar axis. The schematic representation shown in Figure 3a illustrates the polarization bound charges resulting from the head-to-head or tail-to-tail polarization arrangement at domain walls. This type of charge develops as a result of a positive or negative tilt angle of the domain wall with respect to the polar axis. The bound charge density can be estimated as $\rho_{bound} \cong 2P_S \cos(\vec{P}, \vec{n})$[22]. The angle between the domain wall normal $\vec{n}$ and the polarization of the adjacent domain (along z-axis) $\vec{P}$ is extracted from the 3D SHG experiments (Figure 2). This allows us to derive $\rho_{bound}$ and to represent this charge density in a color map superimposed to the surface of the curved wall (Figure 3b).

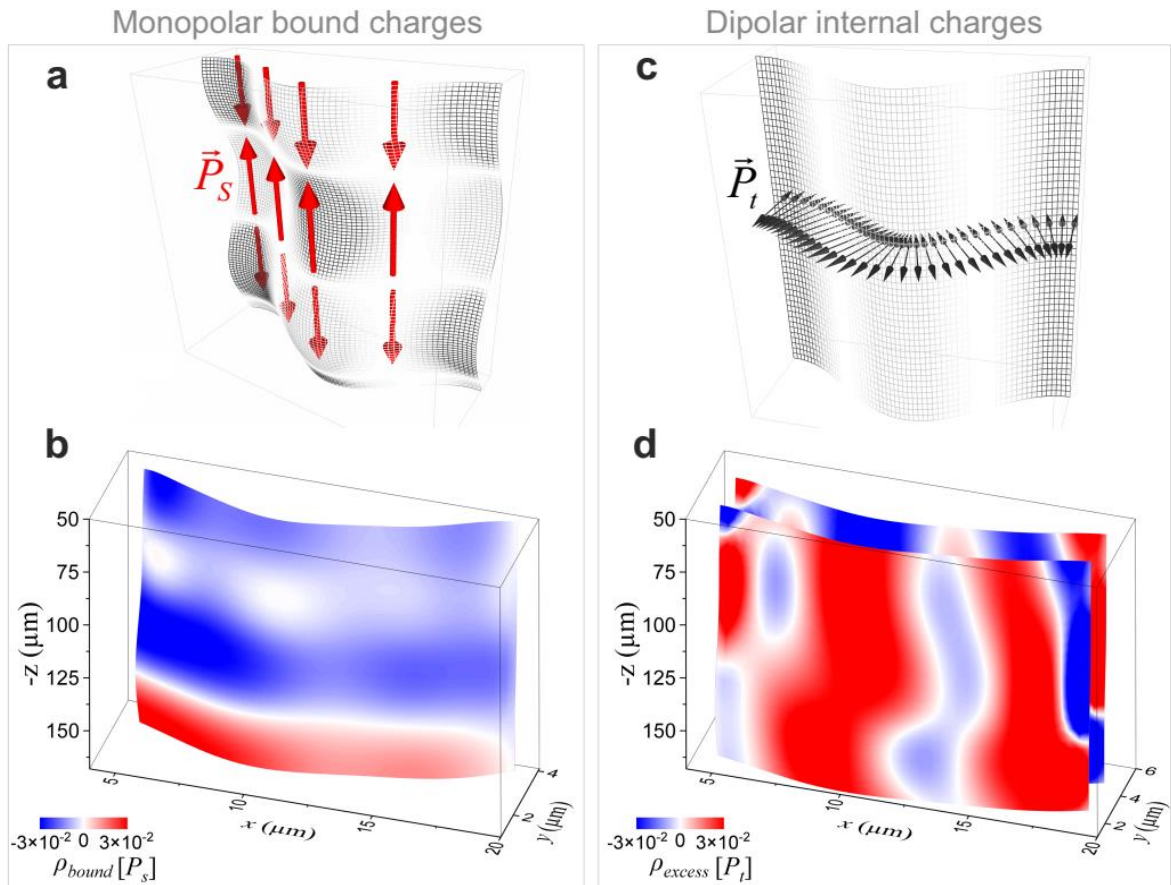

**Figure 3.** (a) Schematic representation showing locally head-to-head or tail-to-tail bound charges due to the local tilting of the domain wall with respect to the polar axis. (b) 2D map of the bound charge density of the curved domain wall in units of the saturation polarization $P_S$. (c) Schematic representation of the internal dipole distribution corresponding to an intrinsically charged Néel-like curved domain wall as viewed by SHG. (d) 2D representation of the variation of the dipole charge density depletion (red) or excess (blue) with the domain wall morphology given in units of the transverse in-plane polarization component $P_t$.

In addition to the dominant bound charges arising from a tilt of the domain wall with respect to the polar axis, complex internal polarization textures and topological structures are also expected to contribute to the electrostatic charge density distribution. The effect of these additional contributions can be particularly perceptible in the case of nominally uncharged domain walls showing a small tilt (Figure S6), and giving rise to a relatively small bound charge density ($\pm$ 5 $10^{-2}$ $P_s$) such as the example considered for this study. In this case, additional contributions can play a role equally important as the monopolar bound charge and may significantly affect the overall charge distribution at the domain wall region. Among possible additional charge contributions, we consider the contribution of the local dipole at intrinsically charged Néel-like domain walls. The curvature of bent domain walls leads in this case to a local excess or depletion of the dipolar charge density depending on the sign of the curvature, as outlined in Figure 3c. Hence, the variation of the relative angle between adjacent dipoles (Figure 2c) is a good quantity to estimate the charge density excess related to the Néel-like internal structure. This additional term is calculated at each site $i$ of the domain wall $\rho_{excess} \cong P_t \sin\left(\theta_{\vec{P}_{i+1}} - \theta_{\vec{P}_i}\right)$ and plotted as a color map in units of the in-plane polarization $P_t$, as displayed in Figure 3d. The monopolar and dipolar bound charge densities seams to both vary with the 3D curvature of the domain wall. Curvature can be defined in different ways to highlight different geometrical aspects such as to describe the degree by which a surface is concave or convex, or whether it is concavely bent in a specific direction but convexly bent in another. To unveil the detailed features in the 3D morphology of the domain wall, we use two different intrinsic measures of the curvature, known as mean curvature ($H$), and Gaussian curvature ($K$), as detailed in Supplementary note S3. Mean curvature expresses the local concavity or convexity of the surface, while Gaussian curvature is sensitive to the presence of saddle- or dome-shaped regions, irrespective of the concave or convex local character of the surface (see Figure S4).

Based on the morphology of the domain wall obtained from 3D SHG microscopy, Gaussian and mean curvatures are derived and displayed in panels a and f of Figure 4. We observe that the electrostatic charge density shows clear correlations with the domain wall curvature. This comparison is automated by calculating the Pearson coefficient $r$ as a quantitative measure of the statistical correlation between the domain wall curvature and the electrostatic charge density based on the covariance method (see Supporting note S4). This coefficient provides information about the correlation strength (no correlation for $r = 0$, and maximum correlation for $r = \pm1$) and direction (correlation for $r > 0$ and anticorrelation for $r < 0$). The variation of the charge density distribution due to the Néel-like character clearly varies with the mean curvature (99% correlations, Figure4d), while the overall charge distribution accounting for the synergetic effect of monopolar and dipolar bound charges correlates with the Gaussian curvature of the domain wall (80% correlations, Figure 4i).

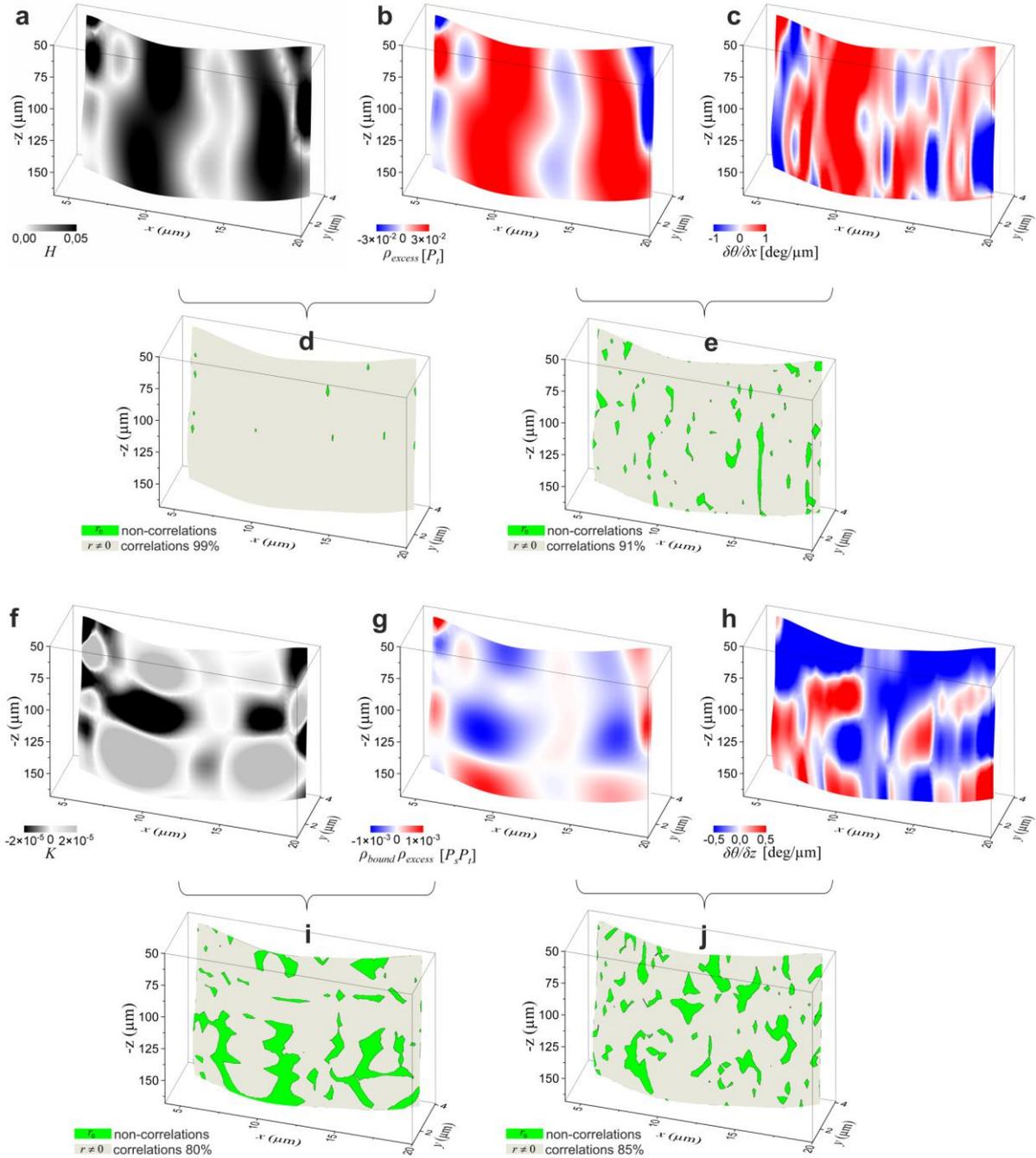

**Figure 4.** (a) Mean curvature $H$ of the domain wall, (b) dipolar charge excess ($\rho_{excess}$) in units of the transverse polarization $P_t$ of the Néel-type domain wall, and (c) variation of the in-plane polarization angle. Local distribution of the statistical correlations (d) between $\rho_{excess}$ and $H$, and (e) $\rho_{excess}$ and the variation of the in-plane polarization angle. (f) Gaussian curvature $K$ of the domain wall, (g) combined monopolar ($\rho_{bound}$) and dipolar ($\rho_{excess}$) charge density in units of $P_S P_t$, and (h) variation of the out-of-plane polarization angle. Local distribution of the statistical correlations (i) between K and the total charge density combining $\rho_{excess}$ and $\rho_{bound}$, and (j) the total charge density and the variation of the out-of-plane polarization angle. In panels d,e,i, and j, the absence of correlations is characterized by a zero Pearson coefficient

0.25<$r_0$<0.25 displayed in green color, while the light grey background represents the absolute value of the non-zero Pearson coefficient.

This result shows that the dependence of the electrostatic charge density on curvature has a direct effect on the orientation of the local polarization angle and its variation throughout the wall profile. Moreover, the domain wall curvature may also affect the electronic conductivity paths linked to the charge excess lines (see, e.g., Figure 4d,g), providing an additional degree of freedom to design 2D resistor networks in the domain wall[54].

Let us now compare the charge density distribution to the internal polarization variation (Figure 4c,h). In the same way as described earlier, this comparison is automated by calculating the Pearson coefficient $r$. The exhaustive statistical correlation results are summarized in Table S1. Best correlations have been observed: (i) between the in-plane variation of the polarization angle and the dipole charge density due to the Néel character, and (ii) between the out-of-plane polarization variation and the joint action of the dipolar and monopolar bound charges (Figure 4e,j).

Analytical calculations in the framework of Landau-Ginzburg-Devonshire (LGD) approach and finite element modeling (FEM) of the polarization distribution in the vicinity of domain walls in uniaxial ferroelectrics (see Methods and problem formulation in the Supplementary note S5), in particular in LiNbO$_3$[22], lead to the (trivial) conclusion that straight and uncharged Ising-type domain walls can be stabilized only in a bulk ferroelectric. This is valid only in the absence of pinning centers (such as lattice barriers, or charged and/or elastic defects), and without significant spatial gradients of elastic, electric and temperature fields. The violation of any of these conditions can lead to a curvature of the domain wall, and very often in these cases the Néel-type polarization appears at the wall due to the omnipresent flexoelectric coupling[23]. In particular, curvature can strongly affect the domain wall structure as well as the domain wall profile (see Figure S7 and the related text). The magnitude of the Néel-like dipolar component can be estimated as $M_{dip} \approx \frac{f}{\sqrt{2}R} P_S^2$, where $f \cong F_{12}\frac{Q_{11}+Q_{12}}{s_{11}+s_{12}}$ represents the effective flexoelectric factor (expressed as a function of the flexoelectric coefficient $F_{12}$, the electrostriction coefficients $Q_{ij}$, and the elastic compliances $s_{ij}$), and $\frac{1}{R} = \frac{1}{R_1} + \frac{1}{R_2}$ is the effective curvature which can be written as a function of the principal curvatures $k_1 = \frac{1}{R_1}$ and $k_2 = \frac{1}{R_2}$. While the monopolar term if of electrostatic origin, and it depends only on the domain wall tilt with respect to the polar axis, the dipolar term is of elastic origin. It is directly related to the flexoelectric effect and it varies with the mean curvature $H = \frac{1}{2}(k_1 + k_2)$ (see Supplementary note S3), as corroborated by the experimental results displayed in panels a,b,d of Figure 4.

In the simplest case of a cylindrical nanodomain, we have $R = r_c$, where $r_c$ is the correlation radius referred to the bulk sample[23] (see details and 3D FEM results in Supporting note S6). The absence of a domain wall tilt with respect to the polar axis leads to a week Néel-type polarization component at the domain wall regions, and closure domains at the top and bottom surfaces (see Figure S8 and Figure S9). The monopolar and dipolar components can have very different magnitudes, and, as a rule, the dipolar component depends on the mean curvature and the correlation radius, while the monopolar component depends on the wall tilt. Yet, in particular cases, in which the domain wall tilt is small such as in the present study, the internal polarization orientation of the domain wall can be affected by the synergetic influence of the domain wall tilting and the intrinsically charged character of the domain wall.

## CONCLUSION

The tomography of a curved domain wall segment was conducted by means of second-harmonic microscopy and its internal polarization structure was derived based on volumetric pixel-by-pixel polarimetry analysis. This allowed us to analyze the impact of curvature on the domain wall's polarization structure. The rotation of the local polarization angle by up to ±15° with respect to the wall normal is explained by polarization reordering due to the local variation of the electrostatic monopolar and dipolar charges with curvature. This work sheds new light onto the intrinsically charged character of Néel-type domain walls and the effect of curvature on the residual dipolar charge, thereby demonstrating that domain wall tilting (resulting in monopolar bound charges) is not the only parameter at play at domain boundaries. These findings pave the way towards the design of energy-friendly nanoelectronic device concepts based on the control of the internal polarization of domain walls and the related functionalities with curvature.

## METHODS

**Poling Procedure.** As a uniaxial ferroelectric crystal, we use the archetypal z-cut congruent lithium niobite crystal. More specifically, a commercially available (Yamaju Ceramics) 200 µm-thick Mg(5%)-doped LiNbO$_3$ is used to engineer a curved domain wall by means of a laser-assisted poling procedure. A HeCd laser (λ = 325 nm, P = 10 µW) was focused through a 10x objective onto the –z-side of the crystal. A voltage of 0.8 kV well below the forward coercive voltage of 1.3 kV was applied parallel to the z-axes, with the polarity chosen to support polarization inversion. Under these conditions, the domain nucleation starts after a about five seconds of exposure to the UV light. We then blocked the UV laser and immediately applied a 1.2 kV bias voltage using liquid electrodes (5wt%-NaCl in deionized H$_2$O). The voltage was then continuously decreased to zero within 2 s. Following this protocol, individual domains of a roundish shapes having mostly straight domain walls with inclination angles of ~ 0.2° can be prepared reproducibly. The scrambled morphology of the domain wall is obtained following the enhancement procedure described in ref.[24]. This produces a bent domain wall with an average tilt angle of 0.3° (see Figure S6) and a local tilt angle up to 1°.

**Second-Harmonic Microscopy and 3D Polarimetry analysis.** The second-harmonic generation measurements are carried out using an inverted scanning confocal microscope in reflection geometry. The fundamental wave is generated with a Spectra Physics Ti:Al$_2$O$_3$ laser (Millenium-Tsunami combination), producing 100 fs pulses with a repetition rate of 80 MHz and a wavelength centered at 800 nm. The laser beam is directed at normal incidence to the sample and focused with a 60x magnification objective lens. The SHG intensity is scanned using computer-controlled stepping motors with 100 nm step size and by recording the SHG signal at each scan step with an exposure time of 20 ms per step, at a power of 20 mW. The output intensity is spectrally filtered and collected into a photomultiplier. Series of SHG images are recorded at different α analyzer angles at a laser polarization along the *x*-axis (φ = 0) or *y*-axis (φ = 90°). Each image of the recorded stack is subdivided into a checkerboard of pixels. An individual polar plot is obtained in each pixel by integrating the local SHG intensity over the stack of images. All the polar plots are then displayed at their pixel center position, yielding a polar plot map (see Figure S1). The local polarization orientation in each pixel is derived by fitting the local polar plots to the analytic form of SHG while accounting for the non-Ising character of the domain wall (see Supplementary note S1). More than 50% of the selected results exhibit a fitting factor $R^2 > 0.96$. The three-dimensional SHG analysis is obtained by repeating the procedure described above at different focal depths, ranging from 50 µm to 170 µm below the surface of the sample.

**Theory and computation Methods.** We conducted finite-element modeling (FEM) based on the Landau-Ginzburg-Devonshire (LGD) theory. FEM is performed using COMSOL@MultiPhysics software, with electrostatics, solid mechanics, and general math (PDE toolbox) modules. The LGD free energy functional of a uniaxial ferroelectric and LGD-type equation are listed in the Supplementary Materials. The ferroelectric, dielectric, and elastic properties of the LiNbO$_3$ core are given in Table S2.

## ASSOCIATED CONTENT

**Supporting Information**

Detailed information on the pixel-by-pixel polarimetry analysis; the determination of the domain wall curvature and polarization profile based on 3D SHG microscopy data; definition of Gaussian and mean curvatures; description of the statistical correlations and summary of the results; theoretical aspects (problem formulation and FEM details) as well as 3D simulation results.

## AUTHOR INFORMATION

**Corresponding Authors**

*Salia Cherifi-Hertel*
Université de Strasbourg, CNRS, Institut de Physique et Chimie des Matériaux de Strasbourg, UMR 7504, 67034, Strasbourg, France ; orcid.org/0000-0002-9617-2098; E-mail: salia.cherifi@ipcms.unistra.fr

*Anna N. Morozovska*
Institute of Physics, National Academy of Sciences of Ukraine, 46, pr. Nauky, 03028 Kyiv, Ukraine; orcid.org/0000-0002-8505-458X; E-mail: anna.n.morozovska@gmail.com

**Authors**

*Ulises Acevedo-Salas, Boris Croes, Yide Zhang, Olivier Cregut, Kokou Dodzi Dorkenoo, Riccardo Hertel*
Université de Strasbourg, CNRS, Institut de Physique et Chimie des Matériaux de Strasbourg, UMR 7504, 67034, Strasbourg, France

*Benjamin Kirbus, Ekta Singh, Henrik Beccard, Michael Rüsing, Lukas M. Eng*
Institute of Applied Physics, Technische Universität Dresden, 01062 Dresden, Germany
ct.qmat: Würzburg-Dresden Cluster of Excellence - EXC 2147, TU Dresden, Germany

*Eugene A. Eliseev*
Institute for Problems of Materials Science, National Academy of Sciences of Ukraine, Krjijanovskogo 3, 03142 Kyiv, Ukraine

**Notes**

The authors declare no competing financial interest.

## ACKNOWLEDGMENTS

This work was supported by the French National Research Agency (ANR) through the "TOPELEC" project (ANR-18-CE92-0052) cofounded by the Deutsche Forschungsgemeinschaft (DFG) (EN-434/41-1). S.C.H. acknowledges the Interdisciplinary Thematic Institute EUR QMat (ANR-17-EURE-0024), as part of the ITI 2021-2028 program supported by the IdEx Unistra (ANR-10-IDEX-0002) and SFRI STRAT'US (ANR-20-SFRI-0012) through the French Programme d'Investissement d'Avenir. A.N.M. acknowledges EOARD project 9IOE063 and related STCU partner project P751. E.S., H.B., M.R., and L.M.E. acknowledge financial support by the DFG within the FOR5044 (ID: 426703838) and the Würzburg-Dresden Cluster of Excellence on "Complexity and Topology in 409 Quantum Matter"- ct.qmat (EXC 2147; ID: 39085490) as well as support by the Light Microscopy Facility, a Core Facility of the CMCB Technology Platform at TU Dresden.

**Supporting Information**

Impact of 3D curvature on the polarization orientation in non-Ising domain walls

*Ulises Acevedo-Salas[1], Boris Croes[1], Yide Zhang[1], Olivier Cregut[1], Kokou Dodzi Dorkenoo[1], Benjamin Kirbus[2], Ekta Singh[2], Henrik Beccard[2], Michael Rüsing[2], Lukas M. Eng[2,3], Riccardo Hertel[1], Eugene A. Eliseev[4], Anna N. Morozovska[5,+], Salia Cherifi-Hertel[1,*]*

[1] Université de Strasbourg, CNRS, Institut de Physique et Chimie des Matériaux de Strasbourg, UMR 7504, 67034, Strasbourg, France

[2] Institute of Applied Physics, Technische Universität Dresden, 01062 Dresden, Germany

[3] ct.qmat: Würzburg-Dresden Cluster of Excellence - EXC 2147, TU Dresden, Germany

[4] Institute for Problems of Materials Science, National Academy of Sciences of Ukraine, Krjijanovskogo 3, 03142 Kyiv, Ukraine

[5] Institute of Physics, National Academy of Sciences of Ukraine, 46, pr. Nauky, 03028 Kyiv, Ukraine

Corresponding authors:
* E-mail: salia.cherifi@ipcms.unistra.fr
+ E-mail: anna.n.morozovska@gmail.com

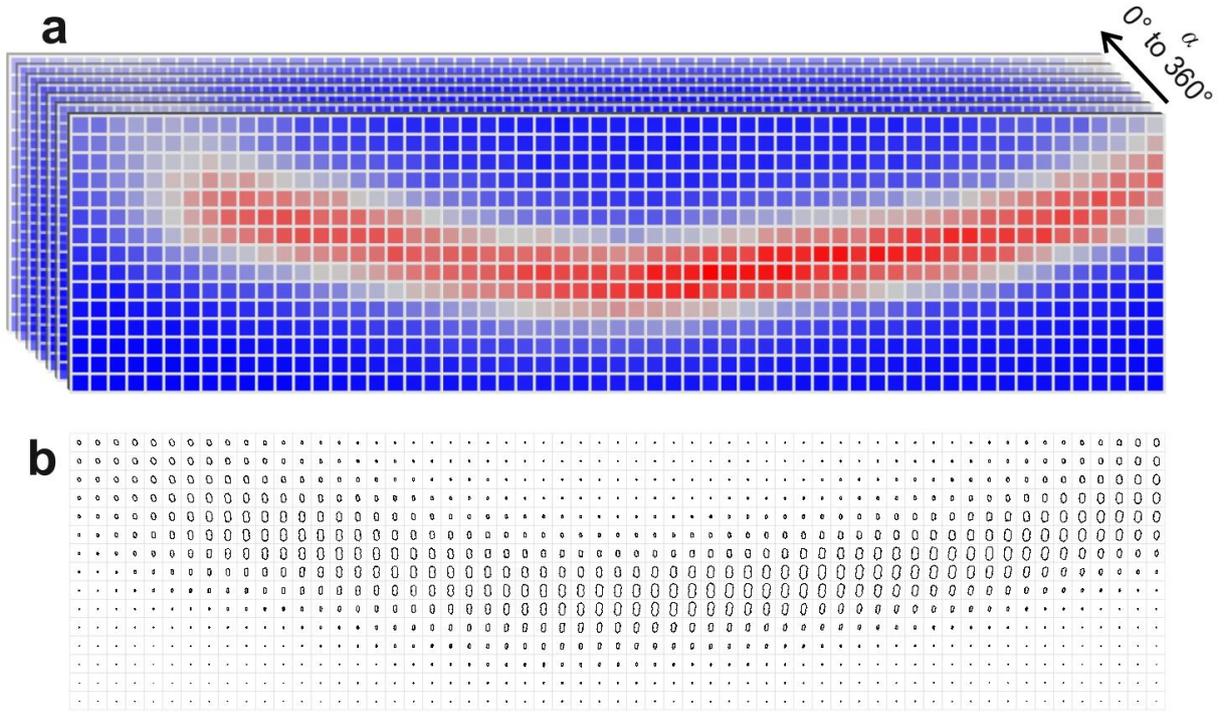

**Figure S1.** (a) Stack of images recorded as a function of the analyzer angle α at a given depth (here, z = 162.5 μm below the top surface). The checkerboard-type pixel-by-pixel resolution is indicated with gray grid lines in (a). The polar-plots representing the variation of the SHG intensity over the stack of images at each pixel are displayed in panel (b).

# S1 Local polarimetry analysis

In the following, we present the general analytic form of the second-harmonic polarimetry accounting for the non-Ising character of ferroelectric domain walls and their respective nonlinear optical susceptibility tensors. The SHG intensity variation with the polarization $\varphi$ of the fundamental wave (FW) and analyzer angle $\alpha$ is given by $I^{SHG}(\varphi, \alpha) = \left|\vec{P}^{2\omega}(\varphi, \alpha)\right|^2$. The tensorial form of the second order polarization induced by the light-matter interaction reads:

$$\begin{pmatrix} P_x(\varphi) \\ P_y(\varphi) \\ P_z(\varphi) \end{pmatrix} = \varepsilon_0 \begin{pmatrix} d_{11} & d_{12} & d_{13} & d_{14} & d_{15} & d_{16} \\ d_{21} & d_{22} & d_{23} & d_{24} & d_{25} & d_{26} \\ d_{31} & d_{32} & d_{33} & d_{34} & d_{35} & d_{36} \end{pmatrix} \begin{pmatrix} E_x^2(\varphi) \\ E_y^2(\varphi) \\ E_z^2(\varphi) \\ 2E_yE_z(\varphi) \\ 2E_zE_x(\varphi) \\ 2E_xE_y(\varphi) \end{pmatrix} \quad (1)$$

where $E_i(\varphi)$ is the electric field of the FW ($E_0\cos\varphi, E_0\sin\varphi, 0$), and $d_{ij}$ represent the elements of the nonlinear optical susceptibility tensor written following the Voigt notation: $2d_{ij} = \chi^{(2)}{}_{ijk}$. The indices $i, j, k$ refer to the Cartesian laboratory coordinates ($x, y, z$). The complete dependence of the SHG response on the analyzer angle $\alpha$ and the input polarization $\varphi$ is obtained using the Jones formalism accounting for the rotation $\alpha$ of a linear polarizer as follows:

$$\begin{pmatrix} P_x(\varphi,\alpha) \\ P_y(\varphi,\alpha) \\ P_z(\varphi,\alpha) \end{pmatrix} = \begin{pmatrix} \cos^2\alpha & \cos\alpha\sin\alpha & 0 \\ \cos\alpha\sin\alpha & \sin^2\alpha & 0 \\ 0 & 0 & 1 \end{pmatrix} \begin{pmatrix} P_x(\varphi) \\ P_y(\varphi) \\ P_z(\varphi) \end{pmatrix} \quad (2)$$

The orientation of the local polarization $P$ is identified with respect to the laboratory coordinates system by the angle $\theta_P$ (taken with respect to the x-axis). The related susceptibility tensor is derived using rotation and transformation matrices and by assuming a Néel or Bloch character of the domain wall, as explained in Refs. [1,2]:

$$\boldsymbol{d}^{Bloch} = \begin{pmatrix} \sin\theta_P & \cos\theta_P & 0 \\ -\cos\theta_P & \sin\theta_P & 0 \\ 0 & 0 & 1 \end{pmatrix} \begin{pmatrix} 0 & 0 & 0 & d_{14} & 0 & d_{16} \\ d_{21} & d_{22} & d_{23} & 0 & d_{25} & 0 \\ 0 & 0 & 0 & d_{34} & 0 & d_{36} \end{pmatrix}$$

$$\times \begin{pmatrix} \sin^2\theta_P & \cos^2\theta_P & 0 & 0 & 0 & -\sin 2\theta_P \\ \cos^2\theta_P & \sin^2\theta_P & 0 & 0 & 0 & \sin 2\theta_P \\ 0 & 0 & 1 & 0 & 0 & 0 \\ 0 & 0 & 0 & \sin\theta_P & \cos\theta_P & 0 \\ 0 & 0 & 0 & -\cos\theta_P & \sin\theta_P & 0 \\ \frac{1}{2}\sin 2\theta_P & -\frac{1}{2}\sin 2\theta_P & 0 & 0 & 0 & -\cos 2\theta_P \end{pmatrix} \quad (3)$$

$$\boldsymbol{d}^{Ne\acute{e}l} = \begin{pmatrix} -\cos\theta_P & \sin\theta_P & 0 \\ -\sin\theta_P & -\cos\theta_P & 0 \\ 0 & 0 & 1 \end{pmatrix} \begin{pmatrix} d_{11} & d_{12} & d_{13} & 0 & d_{15} & 0 \\ 0 & 0 & 0 & d_{24} & 0 & d_{26} \\ d_{31} & d_{32} & d_{33} & 0 & d_{35} & 0 \end{pmatrix}$$

$$\times \begin{pmatrix} \cos^2\theta_P & \sin^2\theta_P & 0 & 0 & 0 & \sin 2\theta_P \\ \sin^2\theta_P & \cos^2\theta_P & 0 & 0 & 0 & -\sin 2\theta_P \\ 0 & 0 & 1 & 0 & 0 & 0 \\ 0 & 0 & 0 & -\cos\theta_P & \sin\theta_P & 0 \\ 0 & 0 & 0 & -\sin\theta_P & -\cos\theta_P & 0 \\ -\frac{1}{2}\sin 2\theta_P & \frac{1}{2}\sin 2\theta_P & 0 & 0 & 0 & \cos 2\theta_P \end{pmatrix} \quad (4)$$

The measurements are conducted by fixing the polarizer angle along the x-axis and rotating the analyzer angle (i.e., $\alpha$-scan at $\varphi = 0$). The obtained results are fitted using the analytic form of the SHG (see above). The fitting function takes the following form for a Néel-type internal domain wall structure:

$$I^{SHG} = \{[A\cos^2\alpha + B\cos\alpha\sin\alpha]^2 + [A\cos\alpha\sin\alpha + B\sin^2\alpha]^2\} + C \quad (5)$$

where C is a constant accounting for a background signal, and the coefficients A and B correspond to the effective nonlinear optical anisotropy factors of the domain wall, defined as:

$$A = -\left(\frac{d_{11}}{d_{12}} - \frac{d_{26}}{d_{12}} - 1\right)\cos^3\theta_P - \left(1 + \frac{d_{26}}{d_{12}}\right)\cos\theta_P \quad (6)$$

$$B = \left(\frac{d_{11}}{d_{12}} - \frac{d_{26}}{d_{12}} - 1\right)\sin^3\theta_P + \left(\frac{d_{26}}{d_{12}} - \frac{d_{11}}{d_{12}}\right)\sin\theta_P \quad (7)$$

# S2 Determination of the domain wall curvature and polarization profile based on the 3D SHG data

The position of the pixels forming the domain wall (red region in Figure S2a) is vectorized, and ($x$, $y$) coordinates are then associated to each pixel for a given measurement depth (z-position), as depicted in Figure S2b. A polynomial function is then used to fit the resulting domain wall geometry (see the red line in Figure S2c). The pixels forming the core of the domain wall are then selected by their proximity to the fitted curve (see square dots in Figure S2d). We use typically 95% confidence bands, as highlighted in light blue color in Figure S2d. The 3D morphology of the domain wall is then reconstructed in the z-direction using a cubic-spline interpolation of the polynomial functions obtained at each z-position (see Figure S2e).

The local polar plots forming the domain wall are fitted using the analytic form of the SHG (given by equation (5)). The local polarization orientation is derived at each pixel as a fit parameter. An average polarization angle $\theta_P$ accounting for the first nearest neighbors is calculated and assigned to the selected pixels (see the blue square in Figure S2d). This procedure is repeated for the data sets measured at each z-position. The 3D distribution of $\theta_P$ is determined by an interpolation of the results in the z-direction.

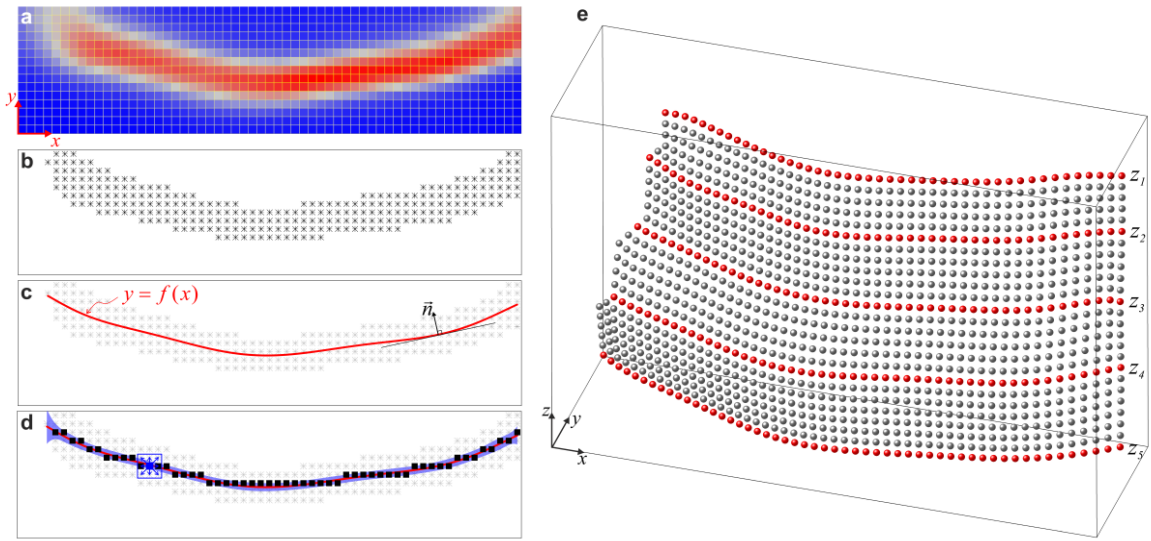

**Figure S2.** 3D Domain wall geometry analysis obtained based on (a) SHG images obtained at a given depth (here, z = 162.5 µm below the surface of the sample). (b) The domain wall region is then vectorized, so that each pixel can be labeled by its ($x$, $y$) coordinates. The red line displayed in panel (c) is a polynomial fit of the central domain wall position $y = f(x)$ in the $xy$-plane. (d) The wide blue line represents a 95% confidence band around the fitted domain wall positions. The 3D reconstruction of the domain wall geometry is obtained by repeating the above mentioned procedure (a-c) at each measurement depth, labeled $z_{1-5}$. (e) The red spheres are discrete points of the fitted polynomial curves at each analyzed depth, while the gray spheres represent the interpolated 3D surface.

# S3  Mean curvature and Gaussian curvature definition

Let us consider a given point $p$ of a regular two-dimensional surface, described in the Euclidean space as shown in Figure S3. The intersection of the normal planes and the surface (depicted in blue in Figure S3) forms curved sections (displayed as dashed lines in Figure S3). Their maximum and minimum curvature is characterized by the principal curvatures $k_1$ and $k_2$, expressed by the eigenvalues of the shape operator at the point $p$. The local mean curvature and Gaussian curvature are defined by the two principal curvatures, $k_1$ and $k_2$, as explained in the following.

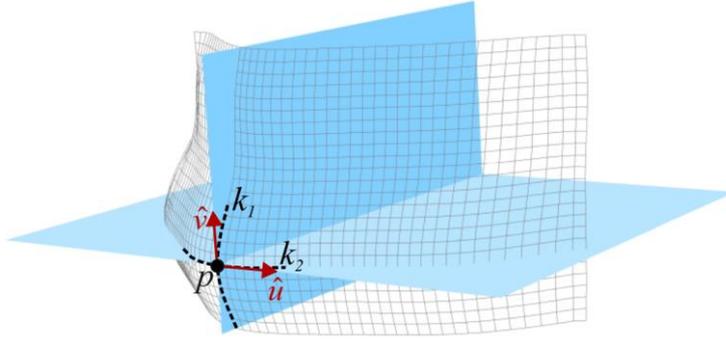

**Figure S3.** Principal curvatures ($k_1$ and $k_2$) and principal directions (given by the unit vectors $\hat{u}$ and $\hat{v}$) at a given point $p$ of the surface.

The mean curvature ($H$) is defined as the mean value of the principal curvatures $k_1$ and $k_2$ (see Supplementary Figure S3)

$$H = \frac{k_1 + k_2}{2} \qquad (8)$$

It can be determined from the first and second fundamental forms of the surface (i.e., the eigenvalues of the shape operator) as:

$$H = \frac{GL - 2FM + EN}{2(EG - F^2)}, \qquad (9)$$

where $E$, $F$ and $G$ are the coefficients of the first, and $L$, $M$ and $N$ are the coefficients of the second fundamental form that describe the full geometry of the surface. These coefficients are numerically derived from discrete points of the reconstructed surface by calculating the corresponding partial derivatives [3-5]:

$$E = \left|\frac{\partial k_1}{\partial u}\right|^2, \quad F = \frac{\partial k_1}{\partial u} \cdot \frac{\partial k_2}{\partial v}, \quad G = \left|\frac{\partial k_2}{\partial v}\right|^2, \qquad \text{Eq (3)}$$
$$L = \sum_i k_i \frac{\partial^2 k_i}{\partial u}, \quad M = \sum_i k_i \frac{\partial^2 k_i}{\partial u \partial v}, \quad N = \sum_i k_i \frac{\partial^2 k_i}{\partial v}. \qquad (10)$$

In the same way, Gaussian curvature ($K$) is defined as the product of the principal curvatures as $K = k_1 k_2$. Alternatively, it can be expressed by the eigenvalues of the shape operator as
$$K = \frac{(LN - M^2)}{(EG - F^2)}. \qquad (11)$$

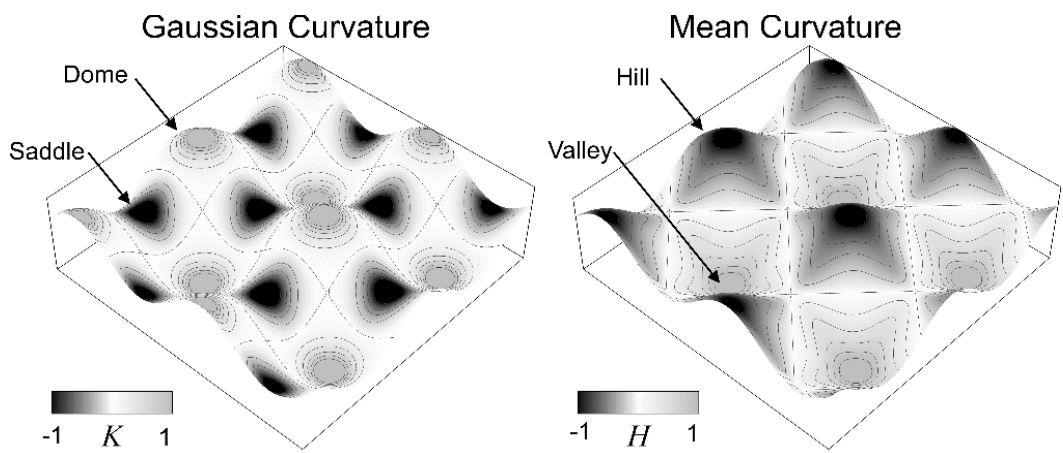

**Figure S4.** Visual comparison of Gaussian curvature ($K$) and mean curvature ($H$).

# S4  Statistical correlations

To establish correlations between parameters A (e.g., $K$ or $H$ curvatures) and B (e.g., $\theta_P$), the data set of $A$ and $B$ is first normalized (intensity ranging from 0 to 255). Then, a sampling grid sub-dividing each dataset into $i \times j$ super-pixels is defined. Each of these super-pixels contains $m \times n$ pixels, where $i$, $m$ are the indexes for the rows, and $j$, $n$ are the indexes for the columns, respectively (see Figure S5). The grid elements of the datasets $A$ and $B$ are then converted into their respective vector representation $a_{ij}$ and $b_{ij}$ containing $N$ elements ($N = m \times n$), as depicted in Figure S5.

The linear relationship between two parameters can be characterized by a correlation coefficient. The Pearson's coefficient $r$ measures the statistical correlation between two continuous variables based on the covariance method. It provides information about the correlation and the direction, i.e., it distinguishes between correlations and anticorrelations, depending on the positive or negative sign of the coefficient.

Pearson's coefficient is determined at each element of $A$ and $B$, by calculating the correlation between the vectors $a_{ij}$ and $b_{ij}$ as:

$$r_{ij}(a_{ij}, b_{ij}) = \frac{1}{N-1} \sum_{k=1}^{N} \left(\frac{a_{ijk} - \mu_a}{\sigma_a}\right)\left(\frac{b_{ijk} - \mu_b}{\sigma_b}\right), \quad (12)$$

where $\mu_a$, $\mu_b$, $\sigma_a$ and $\sigma_b$ are the local mean and the standard deviation of the vectors $a$ and $b$. A correlation value is obtained at each grid element, which is then mapped into a new correlation dataset. The values of $r$ range from -1 to 1, where a value of -1 indicates a perfect negative linear correlation (i.e., anticorrelation), whereas +1 indicates a perfect positive linear correlation. A value of $r = 0$ means that there is no correlation [6,7].

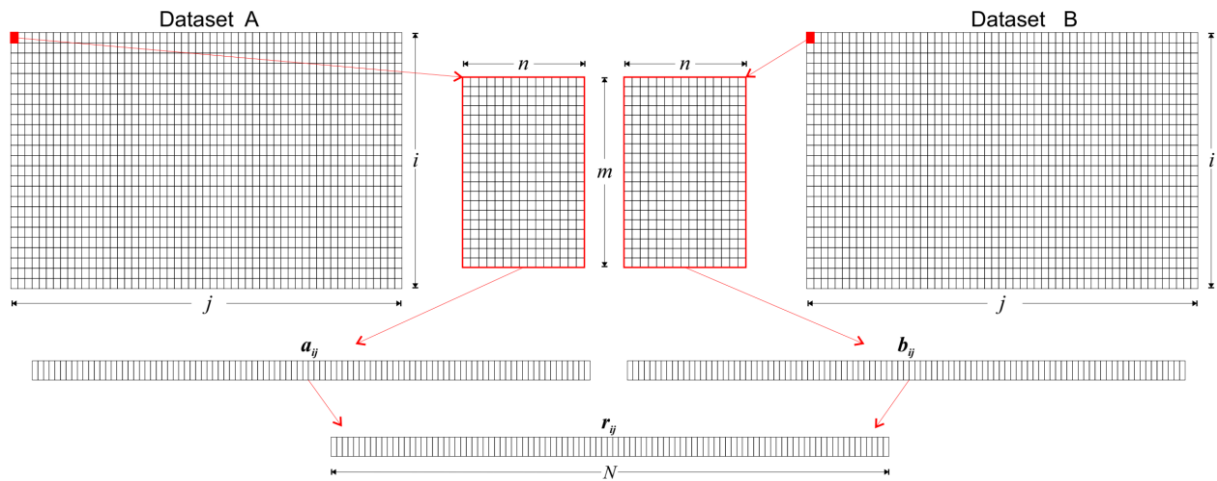

**Figure S5.** Schematic representation of the correlation method between two parameters $A$ and $B$ corresponding in our study to the curvature, $K$ or $H$, and the polarization angle $\theta_P$.

**Table S1.** Summary of the statistical correlations between the domain wall curvature (either $K$ or $H$), the charge density distribution ($\rho_{excess}$ and $\rho_{bound}$) and the variation of the polarization angle. The percentage (%) represents the domain wall surface fraction showing correlations. We denote strong correlations (maximum positive correlation factor) as $r_p$, and strong anticorrelations (maximum negative correlation factor) as $r_n$. Intermediate and zero correlations are labeled $r_i$ and $r_0$, respectively.

| | Strong correlations $r_p \geq 0.75$ $r_n \leq -0.75$ | | Intermediate correlations $0.75 > r_i \geq 0.25$ $-0.75 < r_i \leq -0.25$ | Zero correlations $0.25 > r_0 > -0.25$ |
|---|---|---|---|---|
| $\rho_{excess}$ vs. $H$ | $r_p$ = 1% $r_n$ = 94% | 95 % | 4 % | 1 % |
| $\rho_{excess}.\rho_{bound}$ vs. $H$ | $r_p$ = 47% $r_n$ = 13% | 60 % | 27 % | 13 % |
| $\frac{\delta\theta}{\delta x}$ vs. $\rho_{excess}$ | $r_p$ = 35% $r_n$ = 20% | 55 % | 36 % | 9 % |
| $\frac{\delta\theta}{\delta x}$ vs. $H$ | $r_p$ = 22% $r_n$ = 33% | 55 % | 35 % | 10 % |
| $\theta_P$ vs. $H$ | $r_p$ = 23% $r_n$ = 26% | 49 % | 38 % | 13 % |
| $\rho_{excess}$ vs. $K$ | $r_p$ = 14% $r_n$ = 34% | 48 % | 41 % | 11% |
| $\frac{\delta\theta}{\delta z}$ vs. $H$ | $r_p$ = 24% $r_n$ = 21% | 45 % | 38 % | 17 % |
| $\frac{\delta\theta}{\delta z}$ vs. $\rho_{excess}.\rho_{bound}$ | $r_p$ = 29% $r_n$ = 14% | 43 % | 41 % | 16 % |
| $\frac{\delta\theta}{\delta z}$ vs. $\rho_{bound}$ | $r_p$ = 22% $r_n$ = 16% | 38 % | 40 % | 22 % |
| $\frac{\delta\theta}{\delta x}$ vs. $K$ | $r_p$ = 23% $r_n$ = 14% | 37 % | 42 % | 21 % |
| $\rho_{excess}.\rho_{bound}$ vs. $K$ | $r_p$ = 18% $r_n$ = 16% | 34 % | 46 % | 20 % |
| $\rho_{bound}$ vs. $K$ | $r_p$ = 18% $r_n$ = 16% | 34 % | 43 % | 23 % |
| $\theta_P$ vs. $K$ | $r_p$ = 11% $r_n$ = 22% | 33 % | 44 % | 23 % |
| $\frac{\delta\theta}{\delta z}$ vs. $K$ | $r_p$ = 12% $r_n$ = 20% | 32 % | 47 % | 21 % |
| $\rho_{bound}$ vs. $H$ | $r_p$ = 11% $r_n$ = 12% | 23 % | 43 % | 34 % |

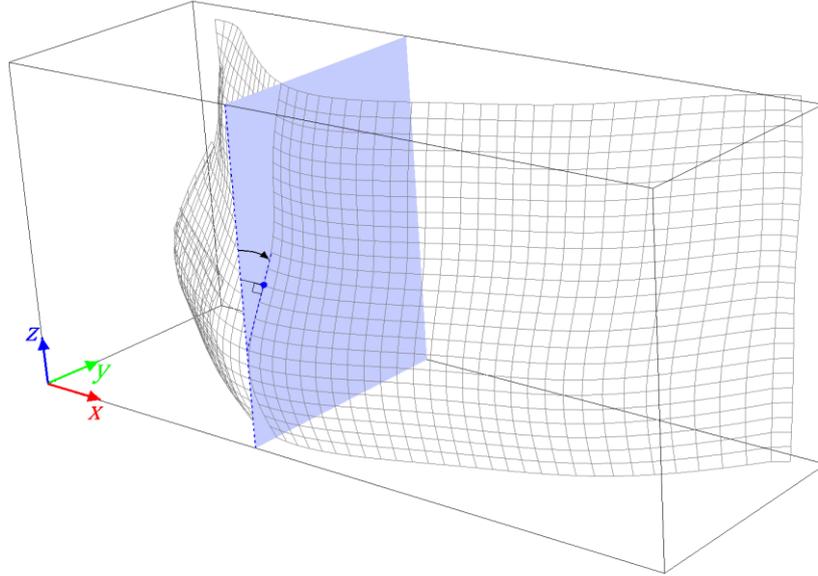

**Figure S6.** Tilt angle of the domain wall as derived from the 3D SHG image. The average value of the angle as derived from the tangent lying in the yz-plane at each point of the domain wall's surface is of 0.3°. The absolute value of the tilt angle varies locally from 0° to 1°.

# S5  Problem formulation and FEM details

The bulk part of the LGD free energy functional $G$ of a uniaxial ferroelectric $LiNbO_3$ includes a Landau energy – an expansion on powers of 2-4-6 of its ferroelectric polarization $P_3$, $G_{Landau}$, a polarization gradient energy, $G_{grad}$; an electrostatic energy, $G_{el}$; an elastic, electrostriction contribution $G_{es}$, and a flexoelectric contribution, $G_{flexo}$. It has the form:

$$G = G_{Landau} + G_{grad} + G_{el} + G_{flexo} + G_{flexo}, \tag{13}$$

$$G_{Landau} = \int_{V_C} d^3r \left[\frac{\alpha}{2}P_3^2 + \frac{\beta}{4}P_3^4 + \frac{\gamma}{6}P_3^6\right], \tag{14}$$

$$G_{grad} = \int_{V_C} d^3r \frac{g_{ij}}{2}\frac{\partial P_3}{\partial x_i}\frac{\partial P_3}{\partial x_j}, \tag{15}$$

$$G_{el} = -\int_{V_C} d^3r \left(P_3 E_3 + \frac{\varepsilon_0 \varepsilon_b}{2}E_i E_i\right), \tag{16}$$

$$G_{es} = -\int_{V_C} d^3r \left(\frac{s_{ijkl}}{2}\sigma_{ij}\sigma_{kl} + Q_{ij33}\sigma_{ij}P_3^2\right), \tag{17}$$

$$G_{flexo} = -\int_{V_C} d^3r \frac{F_{ij3l}}{2}\left(\sigma_{ij}\frac{\partial P_3}{\partial x_l} - P_3\frac{\partial \sigma_{ij}}{\partial x_l}\right). \tag{18}$$

Here $V_C$ is the volume of $LiNbO_3$ sample. The coefficient $\alpha$ linearly depends on temperature $T$, $\alpha(T) = \alpha_T[T - T_C]$, where $\alpha_T$ is the inverse Curie-Weiss constant and $T_C$ is the ferroelectric Curie temperature. The component $\beta$ and $\gamma$ are regarded as temperature-independent. The gradient coefficients $g_{ij}$ are positively defined and regarded as temperature-independent. In Eq.(17), $\sigma_{ij}$ is the stress tensor, $s_{ijkl}$ is the elastic compliances tensor, and $Q_{ijkl}$ is the electrostriction tensor. In the Lifshitz invariant, Eq.(18), $F_{ijkl}$ is the flexoelectric tensor.

In order to find the spatial distribution of the out-of-plane ferroelectric polarization component $P_3$ inside a uniaxial FE and the acting electric field $E_i$ one should solve a coupled problem consisting of the LGD-type equation for $P_3$ and Poisson equation for electric potential $\phi$:

$$[\alpha_T(T-T_C) - Q_{ij33}\sigma_{ij}]P_3 + \beta P_3^3 + \gamma P_3^5 - g_{11}\frac{\partial^2 P_3}{\partial z^2} - g_{44}\left(\frac{\partial^2 P_3}{\partial x^2} + \frac{\partial^2 P_3}{\partial y^2}\right) = E_3 - F_{ijk3}\frac{\partial \sigma_{ij}}{\partial x_k}. \quad (19)$$

$$\varepsilon_0\varepsilon_b\left(\frac{\partial^2}{\partial x^2} + \frac{\partial^2}{\partial y^2} + \frac{\partial^2}{\partial z^2}\right)\phi = \frac{\partial P_3}{\partial z} - e(Z_d N_d^+ - n), \quad (20)$$

Here $\varepsilon_0$ is the universal electric constant, $\varepsilon_b$ is a background permittivity [8]. There may be free charge is LNO. To account for it, the concentration of the ionized donors, $N_d^+$, and free electrons, $n$, inside the semiconducting channel, which depend on the electric potential $\phi$ in the conventional way [9]:

$$N_d^+(\phi) = N_d^0 f\left(\frac{-E_d + eZ_d\phi + E_f}{k_B T}\right), \quad (21)$$

$$n(\phi) = N_C F_{1/2}\left(\frac{e\phi + E_f - E_C}{k_B T}\right) \quad (22)$$

Here $E_d$ is the donor energy levels respectively, $T$ is the absolute temperature, $k_B$ is a Boltzmann constant, $E_C$ is the position of conduction band bottom. The effective density of states in the conductive band $N_C = \left(\frac{m_n k_B T}{2\pi\hbar^2}\right)^{3/2}$, $m_n$ is the effective mass of electrons. $E_F$ is the Fermi energy level in equilibrium, which is determined from the electroneutrality condition $Z_d N_d^+ - n = 0$ at zero potential. In Eq.(21-22) we introduced the Fermi-Dirac distribution function and the Fermi integral, $f(\xi) = \frac{1}{1+\exp(\xi)}$ and $F_{1/2}(\xi) = \frac{2}{\sqrt{\pi}}\int_0^\infty \frac{\sqrt{\zeta}d\zeta}{1+exp(\zeta-\xi)}$, respectively.

The elastic equations of state for the strains $u_{ij}$ follow from the variation of the LGD free energy with respect to elastic stresses $\sigma_{ij}$:

$$u_{ij} = s_{ijkl}\sigma_{kl} + \beta_{ij}(T-T_0) + F_{ijkl}\frac{\partial P_l}{\partial x_k} + Q_{ijkl}P_k P_l, \quad 0 \leq z \leq h. \quad (23)$$

Equations (23) should be solved along with equations of mechanical equilibrium
$$\partial \sigma_{ij}(\mathbf{x})/\partial x_i = 0, \quad (24),$$
and compatibility equations, $e_{ikl}e_{jmn}\partial^2 u_{ln}(\mathbf{x})/\partial x_k \partial x_m = 0$, which are equivalent to the continuity of elastic displacements $U_i$ [10].

The finite element modeling (FEM) is performed in a COMSOL@MultiPhysics software, using electrostatics, solid mechanics, and general math (PDE toolbox) modules. The ferroelectric, dielectric, and elastic properties of the LiNbO$_3$ core are given in Table S2.

**Table S2.** LGD coefficients and other material parameters of a bulk LiNbO$_3$

| Parameter | Dimension | Values for LiNbO$_3$ collected from Refs.[11-16] |
|---|---|---|
| $\varepsilon_b$ | 1 | 4.6 [9] |
| $\alpha_T$ | m/(F K) | 1.569×10$^6$ [10] |
| $T_C$ | K | 1477 Ref. [11] |
| $\beta$ | C$^{-4}$·m$^5$J | 2.31×10$^{9*}$ |
| $\gamma$ | C$^{-6}$·m$^9$J | 1.76×10$^{9*}$ |
| $g_{ij}$ | m$^3$/F | $g_{44}$=7.96×10$^{-11}$ ** [12] |
| $s_{ij}$ | 1/Pa | $s_{11}$ =5.78×10$^{-12}$, $s_{12}$= −1.01×10$^{-12}$, $s_{13}$= −1.47×10$^{-12}$, $s_{33}$= +5.02×10$^{-12}$, $s_{14}$= −1.02×10$^{-12}$, $s_{44}$= 17.10×10$^{-12}$ [13] |
| $Q_{ij}$ | m$^4$/C$^2$ | $Q_{33}$= +0.016, $Q_{13}$= −0.003 |
| $F_{ij}$ | m$^3$/C | $F_{11}$=1.0×10$^{-11}$, $F_{12}$=0.9×10$^{-11}$, $F_{44}$=3×10$^{-11}$ (estimate) |

* The estimation is based on the values of the spontaneous polarization and permittivity at room temperature
** the order of magnitude is estimated from the uncharged domain wall width [12].

To illustrate the effect of curvature on the polarization profile across the curved domain wall, and to take into account all above effects in a self-consistent way, we performed the FEM (see Figure S7). We consider the situation when the curved wall separates a nanodomain induced by the electric field of a biased heated tip, which is placed near the unscreened surface of LiNbO$_3$. The inhomogeneous laser heating of the LiNbO$_3$ can stabilize the curved wall after the electric bias is switched off. We observed the evident correlation between the wall width, its bound charge and its curvature. As expected, the correlation is strongest under the absence of free screening charges. Namely, the strongly curved and charged part of wall is thick, the slightly curved part is thinner, and the straight uncharged domain wall is the thinnest, respectively. Moreover, we see that the strong flexoelectric coupling and high concentration of the screening charges significantly help the wall to maintain the curvature without making the wall thicker. So, it may seem that the curvature does not to act noticeably on the wall structure. However, it is not exactly so, because the screening by free charges compensates the monopolar and dipolar components of the bound charge, while the flexoelectric coupling defines the Neel-type dipolar component only and changes the wall elastic energy. The wall width correlates with its curvature, and the correlation is the strongest under the absence of screening charges (see **Fig. S7**). Plots **S7a, b, c** illustrate the shape of stable domains calculated by FEM for increasing tip voltage. Plots **S7d, e, f** show zoomed portions of the strongly curved and thick, slightly curved and thinner, and the thinnest straight domain walls, respectively.

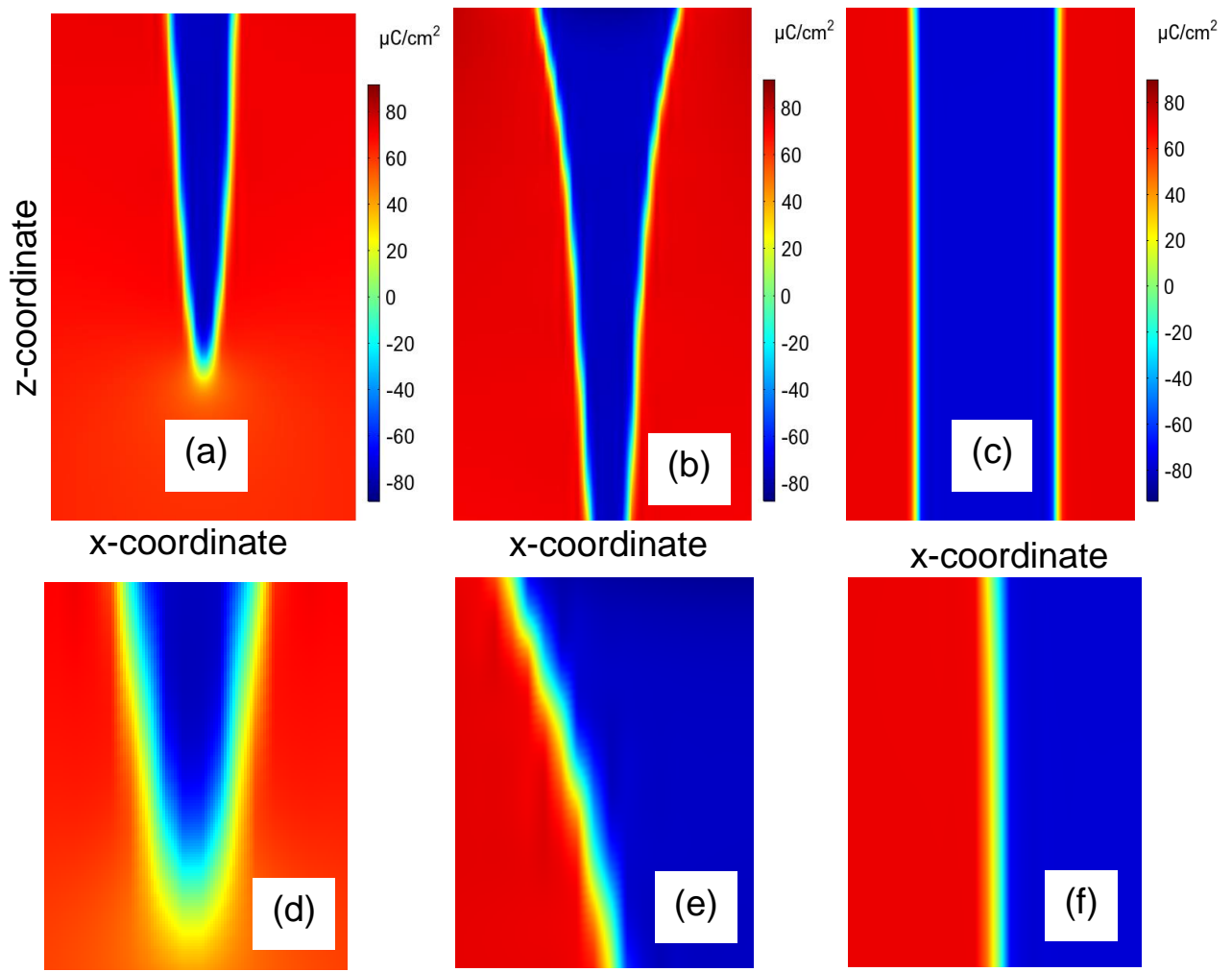

**FIG. S7**. Spatial distributions ($xz$ cross-sections) of the polarization $P_3$ in a thick LiNbO$_3$ layer. Plots **(a, b, c)** illustrate the shape of stable domains calculated by FEM for increasing tip voltage. Plots **(d, e, f)** show zoomed pieces of the strongly curved, slightly curved and straight domain walls, respectively. Before the domain formation the layer was homogeneously polarized. Material parameters are listed in **Table SI,** $T =$293 K.

# S6  3D polarization profile of a non-tilted nano-cylinder derived from FEM calculations

The total free energy density reads as:

$$G = \Delta G_b + \Delta G_{elast} + \Delta G_{strict} + \Delta G_{inhomo} - P_i E_i^d / 2 \tag{25}$$

The polarization-dependent density $\Delta G_b$ can be written as a Taylor series expansion of the polarization components $P_i$ as:

$$\Delta G_b = a_1(P_1^2 + P_2^2) + a_3 P_3^2 + a_{11}(P_1^2 + P_2^2)^2 + a_{13}(P_1^2 + P_2^2)P_3^2 + a_{15}P_1(P_1^2 - 3P_2^2)P_3 + a_{33}P_3^4 + a_{333}P_3^6 \tag{26}$$

The electrostriction contribution to the free energy is

$$\Delta G_{strict} = -Q_{11}(\sigma_1 P_1^2 + \sigma_2 P_2^2) - Q_{33}\sigma_3 P_3^2 - Q_{12}[\sigma_1 P_2^2 + \sigma_2 P_1^2] - Q_{13}(P_1^2 + P_2^2)\sigma_3 - Q_{31}P_3^2(\sigma_1 + \sigma_2) - Q_{44}(\sigma_4 P_2 P_3 + \sigma_5 P_3 P_1) - Q_{51}P_3(P_1(\sigma_1 - \sigma_2) - 2P_2\sigma_6) - Q_{66}P_1 P_2 \sigma_6 \quad (27),$$

and the elastic energy is

$$\Delta G_{elast} = -\frac{1}{2}s_{11}(\sigma_1^2 + \sigma_2^2) - \frac{1}{2}s_{33}\sigma_3^2 - s_{12}\sigma_1\sigma_2 - s_{13}(\sigma_2\sigma_3 + \sigma_3\sigma_1) - \frac{1}{2}s_{44}(\sigma_4^2 + \sigma_5^2) - s_{15}\big((\sigma_1 - \sigma_{22})\sigma_5 - 2\sigma_4\sigma_6\big) - \frac{1}{2}s_{66}\sigma_6^2 \tag{28}$$

The polarization gradient contribution to free energy (Ginzburg term) is

$$\Delta G_{inhomo} = \frac{g_{11}}{2}\left(\left(\frac{\partial P_1}{\partial x_1}\right)^2 + \left(\frac{\partial P_2}{\partial x_2}\right)^2 + \left(\frac{\partial P_3}{\partial x_3}\right)^2\right) + g_{12}\left(\frac{\partial P_1}{\partial x_1}\frac{\partial P_2}{\partial x_2} + \frac{\partial P_1}{\partial x_1}\frac{\partial P_3}{\partial x_3} + \frac{\partial P_3}{\partial x_3}\frac{\partial P_2}{\partial x_2}\right) + \frac{g_{44}}{2}\left(\left(\frac{\partial P_1}{\partial x_2} + \frac{\partial P_2}{\partial x_1}\right)^2 + \left(\frac{\partial P_2}{\partial x_3} + \frac{\partial P_3}{\partial x_2}\right)^2 + \left(\frac{\partial P_3}{\partial x_1} + \frac{\partial P_1}{\partial x_3}\right)^2\right) + g_{15}\left(\left\{\frac{\partial P_1}{\partial x_1} - \frac{\partial P_2}{\partial x_2}\right\}\frac{\partial P_3}{\partial x_1} - \left(\frac{\partial P_1}{\partial x_2} + \frac{\partial P_2}{\partial x_1}\right)\frac{\partial P_3}{\partial x_2}\right) \tag{29}$$

**Table S3.** Polarization expansion coefficients of LiNbO3 used in these simulations.

| parameter | $a_1$ | $a_{3T}$ | $T_0$ | $a_{11}$ | $a_{12}$ | $a_{15}$ | $a_{13}$ | $a_{33}$ | $a_{333}$ | $g_{11}$ | $g_{12}$ | $g_{44}$ | $g_{15}$ |
|---|---|---|---|---|---|---|---|---|---|---|---|---|---|
| value | 6.725 10$^8$ | 7.845 10$^5$ | 1477 | 9 10$^{10}$ | 2$a_{11}$ | 5 10$^9$ | $a_{12}$ | 5.775 10$^8$ | 2.933 10$^8$ | 5 10$^{-10}$ | -0.2 10$^{-10}$ | 0.796 10$^{-10}$ | 0.0 |
| units | $\frac{m}{F}$ | $\frac{m}{F\,K}$ | K | $\frac{m^5}{F\,C^2}$ | $\frac{m^5}{F\,C^2}$ | $\frac{m^5}{F\,C^2}$ | $\frac{m^5}{F\,C^2}$ | $\frac{m^5}{F\,C^2}$ | $\frac{m^9}{F\,C^4}$ | $\frac{m^3}{F}$ | $\frac{m^3}{F}$ | $\frac{m^3}{F}$ | $\frac{m^3}{F}$ |

**Table S4.** Electrostriction coefficients and elastic stiffness coefficients of LiNbO3 used in these simulations.

| parameter | $Q_{11}$ | $Q_{12}$ | $Q_{13}$ | $Q_{31}$ | $Q_{33}$ | $Q_{44}$ | $Q_{51}$ | $Q_{66}$ | $c_{11}$ | $c_{12}$ | $c_{13}$ | $c_{15}$ | $c_{33}$ | $c_{44}$ | $c_{66}$ |
|---|---|---|---|---|---|---|---|---|---|---|---|---|---|---|---|
| value | 0.03 | -0.01 | -5 10$^{-3}$ | -3 10$^{-3}$ | 16 10$^{-3}$ | 0.04 | -0.03 | 2($Q_{11}$ - $Q_{12}$) | 19.89 | 5.467 | 6.726 | 0.783 | 23.37 | 5.985 | $\frac{c_{11} - c_{12}}{2}$ |
| units | $\frac{m^4}{C^2}$ | $\frac{m^4}{C^2}$ | $\frac{m^4}{C^2}$ | $\frac{m^4}{C^2}$ | $\frac{m^4}{C^2}$ | $\frac{m^4}{C^2}$ | $\frac{m^4}{C^2}$ | | 10$^{10}$Pa | 10$^{10}$Pa | 10$^{10}$Pa | 10$^{10}$Pa | 10$^{10}$Pa | 10$^{10}$Pa | |

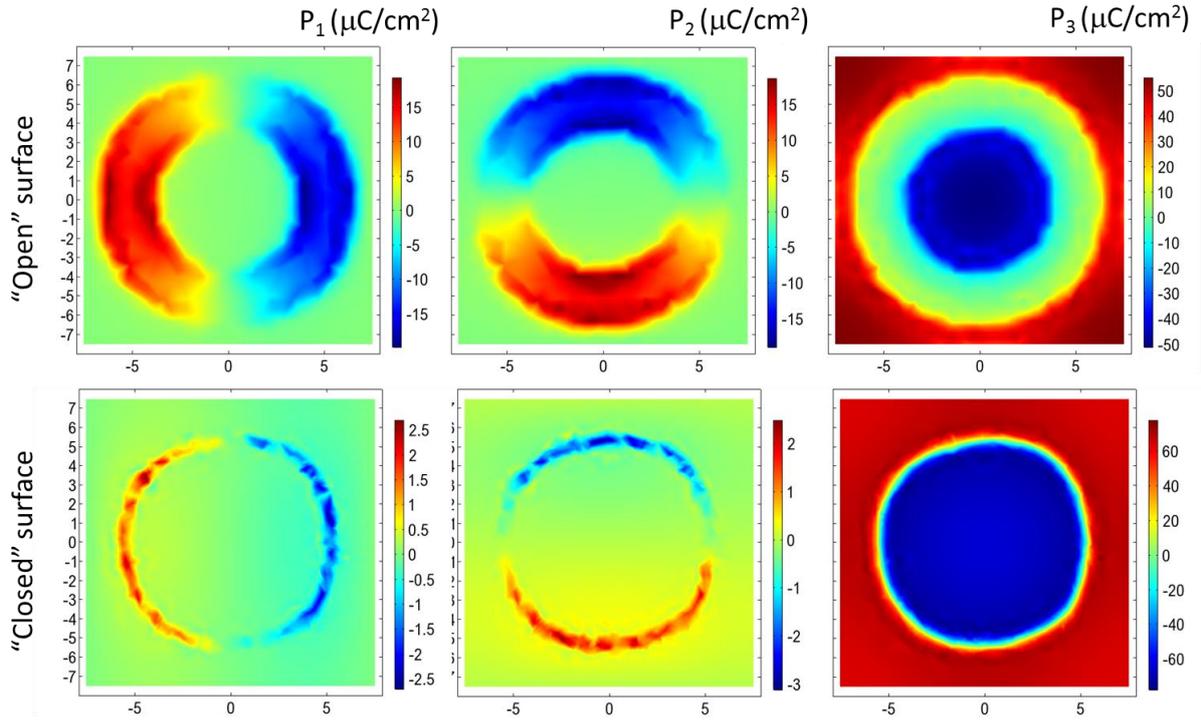

**FIG. S8**. Color map of the three polarization components $P_1$, $P_2$ and $P_3$ in a non-tilted cylindrical-shape domain. The FEM calculations where conducted using a tetrahedral mesh for both "Open" (top row) and "closed" (bottom row) boundary conditions. Clear in-plane polarization is observed around the cylindrical-shape domain with more or less broadening, depending on the chosen boundary conditions.

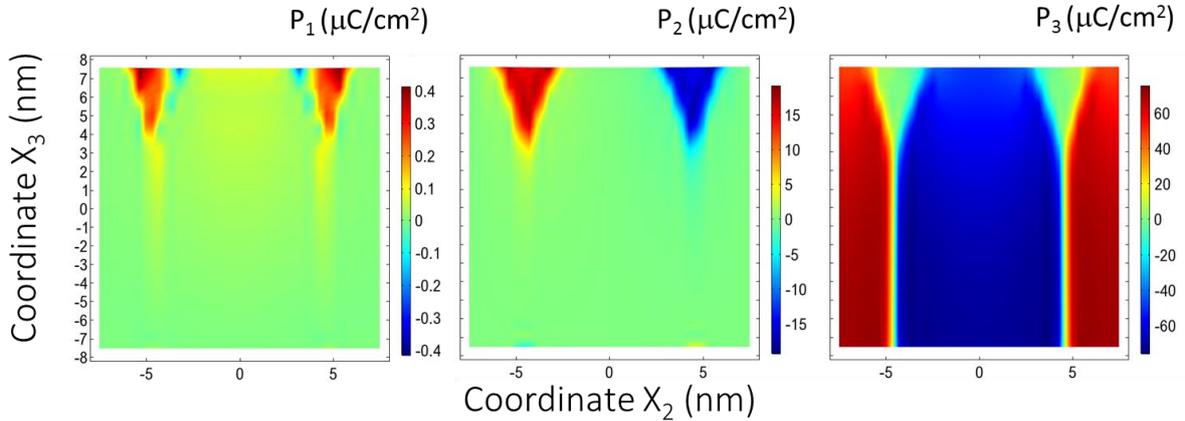

**FIG. S9**. Distribution of the in-plane polarization components $P_1$, $P_2$ and the out-of-plane polarization $P_3$ in a vertical cross-section. Closure domains are observed at the near "open" surface and a weaker in-plane polarization is observed at the domain wall regions. This distribution is almost independent of the anisotropy of the LGD coefficients and the mesh type, but it strongly depends on the surface screening length $\lambda$. For instance, the closure domains practically vanish in the limit $\lambda \rightarrow 0$ ("closed" surface).

# Supplementary References